\documentclass[12pt]{iopart}

\usepackage{iopams}
\usepackage{graphicx}
\begin{document}
\title{Thermal relaxation of magnetic clusters in amorphous Hf$_{57}$Fe$_{43}$ alloy}
\author{Damir Paji\'{c}$^{1}$\footnote{Corresponding author, tel:+385 1 4605555, fax:+385 1 4680336},  Kre\v{s}o Zadro$^{1}$, Ramir Risti\'{c}$^{2}$, Ivica \v{Z}ivkovi\'{c}$^{3}$, \v{Z}eljko Skoko$^{1}$ and Emil Babi\'{c}$^{1}$}
\address{$^{1}$ Department of Physics, Faculty of Science, University of Zagreb,
Bijeni\v{c}ka c. 32, HR-10000 Zagreb, Croatia}
\address{$^{2}$ Department of Physics, University of Osijek, Trg Ljudevita Gaja 6, HR-31000 Osijek, Croatia}
\address{$^{3}$ Institute of Physics, Bijeni\v{c}ka c. 46, HR-10000 Zagreb, Croatia}

\ead{dpajic@phy.hr}
\begin{abstract}
The magnetization processes in binary magnetic/nonmagnetic amorphous alloy Hf$_{57}$Fe$_{43}$ are investigated by the detailed measurements of magnetic hysteresis loops, temperature dependence of magnetization, relaxation of magnetization and magnetic ac susceptibility, including a nonlinear term. Blocking of magnetic moments at lower temperatures is accompanied with the slow relaxation of magnetization and magnetic hysteresis loops. All of the observed properties are explained with the superparamagnetic behaviour of the single domain magnetic clusters inside the nonmagnetic host, their blocking by the anisotropy barriers and thermal fluctuation over the barriers accompanied by relaxation of magnetization. From magnetic viscosity analysis based on thermal relaxation over the anisotropy barriers it is found out that magnetic clusters occupy the characteristic volume from 25 up to 200 nm$^{3}$. The validity of the superparamagnetic model of Hf$_{57}$Fe$_{43}$ is based on the concentration of iron in the Hf$_{100-x}$Fe$_{x}$ system that is just below the threshold for the long range magnetic ordering. This work throws more light on magnetic behaviour of other amorphous alloys, too.
\end{abstract}
\pacs{75.20.-g, 75.50.Bb, 75.50.Kj, 76.60.Es}
%
\submitto{\JPCM}
%
%
\section{Introduction}
\label{introduction}

Magnetism in nanostructured materials has been a very popular topic and a subject of intense research for many years \cite{europhysnews}. Concerning the magnetic colloid \cite{colloid} the nanometre sized magnetic objects have been the subject of interest for quite some time. Now, a large impact of the nanoparticulated and nanostructured magnetics can be seen in many products, as well as in high-tech devices. Besides commercial purposes, the nanostructured magnetic materials pose a broad spectrum of the fundamental physical phenomena, which became accessible with development of the synthesis and characterisation of nanomagnets.

When talking about the magnetic particles of size below about 100nm, superparamagnetism \cite{superparamagnetism} is an usual keyword. It has been shown theoretically that for the particles of these sizes it is favourable to be singledomain \cite{singledomain}, as was measured in magnetic colloid \cite{colloidmeasure}, too. The giant magnetic moments of these particles fluctuate over the anisotropy barrier according to the activation law \cite{fluctdom}. At low temperatures this fluctuation becomes slower than the measurement of one point resulting with very rich phenomenology of non-equilibrium systems and slow relaxation of their magnetization. 

In magnetic alloys it is possible that the (ferro)magnetic single-domain clusters within a non-magnetic or a much weaker magnetic matrix are formed \cite{crangle}. The alloy exhibits a long range ferromagnetic ordering for high concentration of the magnetic atoms, whereas for low magnetic atom contentrations it exhibits paramagnetic behaviour, and somewhere between it is superparamagnetic. The modelling and computational simulation of processes in alloys is still very actual and the experimental magnetic results are reproduced very well \cite{alloycompute}, which helps to establish a connection between microscopic picture and macroscopic properties. 

Magnetic ordering in the binary Hf$_{100-x}$Fe$_{x}$ system was studied previously for different iron concentration $x$ \cite{HfFe,HfFeRistic}. For $x \ge 50$ the long range magnetic order was observed \cite{HfFe} with critical temperatures up to 300K. For $x \le 40$ the system is paramagnetic with increasing of the Curie's constant as $x$ increases \cite{HfFeRistic}. After the observation of splitting between the zero-field-cooled and field-cooled magnetization curves and slow relaxation of magnetization together with hysteresis curves at low temperatures \cite{HfFePajic}, the detailed magnetic investigation of Hf$_{57}$Fe$_{43}$ amorphous alloy has been undertaken. In this paper the results will be presented and explained within the framework of superparamagnetism of magnetic clusters, their blocking by the anisotropy barriers and thermal fluctuation over the barriers accompanied by relaxation of magnetization. 

\section{Experimental procedure}
\label{experimental}

The investigated binary magnetic/nonmagnetic amorphous alloy Hf$_{57}$Fe$_{43}$ within the Hf-Fe system was prepared using melt-spinning method. Starting elements were precisely weighted to fulfil the desired molar proportion and after the whole process total mass did not change. Melting in the argon atmosphere was performed more times in order to obtain compositionally homogeneous alloy.

X-ray diffraction (XRD) patterns were taken at room temperature using an automatic Philips diffractometer, model PW1820 (Cu-K$\alpha $ radiation, graphite monochromator, proportional counter), in Bragg-Brentano geometry. The diffraction intensity was measured in the angular range $20^{\circ } \le 2\theta \le 80^{\circ }$.

Magnetic measurements were performed using a Quantum Design MPMS5 SQUID magnetometer, which uses the extraction method to measure the magnetic moment of the sample with a very high accuracy. Due to the high stability of temperature and stable homogeneous magnetic field, this equipment is very suitable for long-lasting magnetic relaxation measurements. 

The dependence of magnetic moment of the sample $m$ on the temperature $T$ is measured using two modes: zero-field-cooled (ZFC) and field-cooled (FC), both of them during the increase of $T$. The temperature below which the splitting of ZFC and FC curves appears is called blocking temperature $T_{B}$. It is important to mention that there are two different definitions of $T_{B}$ with slightly different values: whether the temperature at which the ZFC curve attains a maximum ($T_{max}$), or the temperature below which the splitting appears ($T_{irr}$), both having a reasonable interpretation. 

The $m(H)$ curves for the applied magnetic field $\mu _{0}H$ up to 5.5T at different stable temperatures were measured. Hysteresis loops were measured with maximum applied field 0.2T because all the curves are reversible above this field. 

Very detailed and precise measurement of the relaxation of magnetic moment of the sample at broad range of stable temperatures from 1.8K up to 25K was performed. The sample was at first heated to 100K that is well-above $T_{B}$ in zero applied magnetic field. Then the magnetic field of 0.01T was imposed. After some waiting time the sample was cooled down to the desired temperature and stabilised. Finally, the magnetic field was reversed to the opposite direction (from 0.01T to -0.01T) and $m$ was measured as the time elapsed during $\sim $3 hours. This procedure was repeated for many different target temperatures below $T_{B}$ with very high reproducibility.

AC susceptibility measurements were performed using the commercial CryoBIND system. The first and the third harmonic were measured simultaneously with two lock-in amplifiers connected in parallel. The amplitude and the frequency of the driving field were set to 1mT and 990Hz, respectively.

\section{Results of measurements}\label{results}

\begin{figure}[t]
\centerline{\includegraphics[width=9cm,clip]{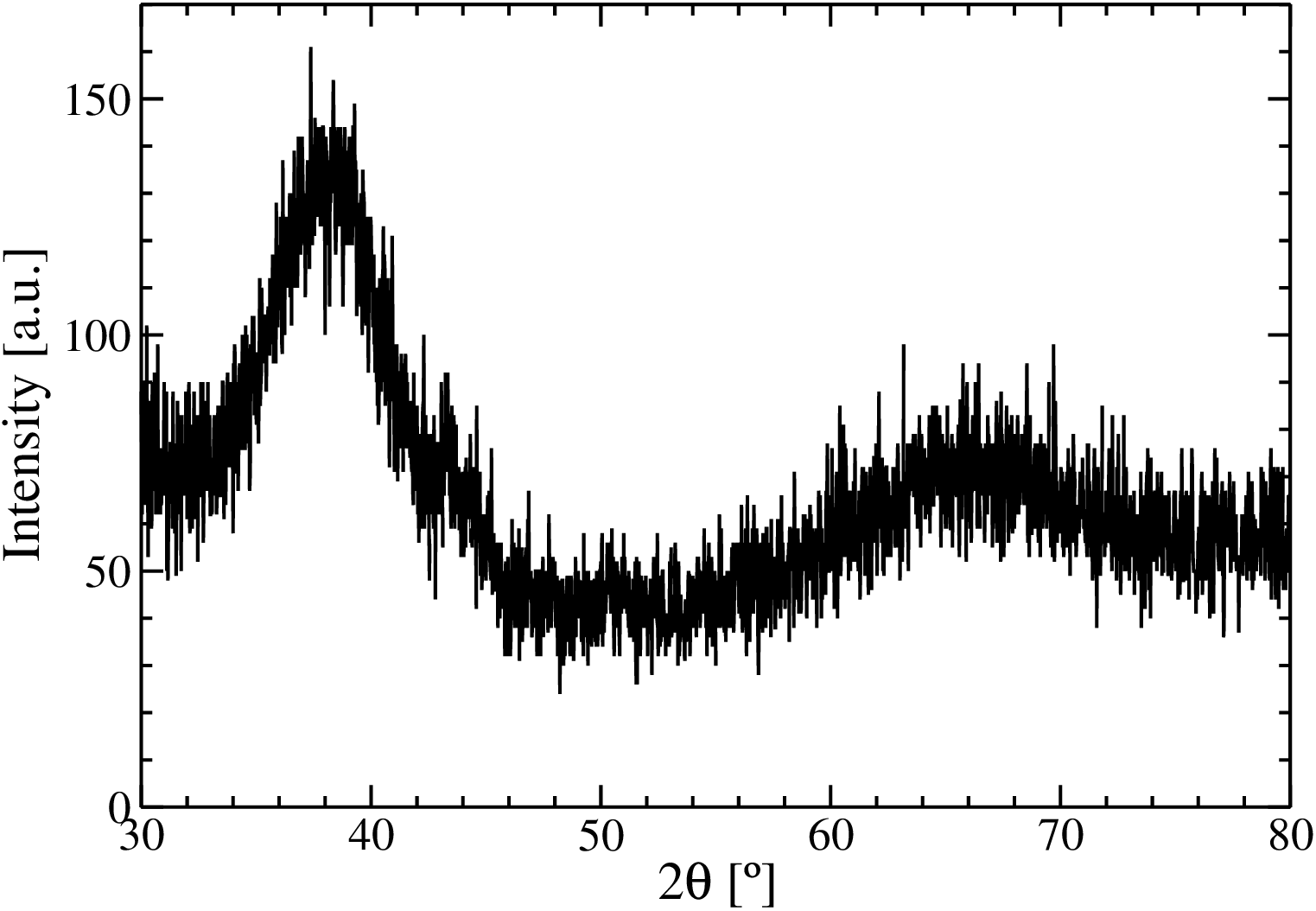}}
\caption{XRD pattern of the as-quenched Hf$_{57}$Fe$_{43}$ sample.}
\label{xrd}
\end{figure}

In order to check the structure, the XRD patterns of the as-quenched Hf$_{57}$Fe$_{43}$ samples were taken and the result is shown in \fref{xrd}. No crystallite peak was detected, but two extremely broad maxima centered at 2$\theta $ of 39$^{\circ }$ and 66$^{\circ }$ were observed, thus indicating that the alloy was in a completely amorphous state. The XRD measurements were done exposing both as-quenched ribbon surfaces to X-ray beam, the surface of the ribbon which was in contact with the inert atmosphere and the surface which was in contact with the wheel. It was done to find out whether there was any difference in the crystal structure between the two surfaces, which could be due to different treatment applied to each surface. The XRD patterns were similar therefore showing the amorphous nature of both sides of the ribbon. Positions of these broad maxima correspond to the results presented in \cite{xrdhffe} for a similar Hf-Fe system.

\begin{figure}[b]
\centerline{\includegraphics[width=9cm,clip]{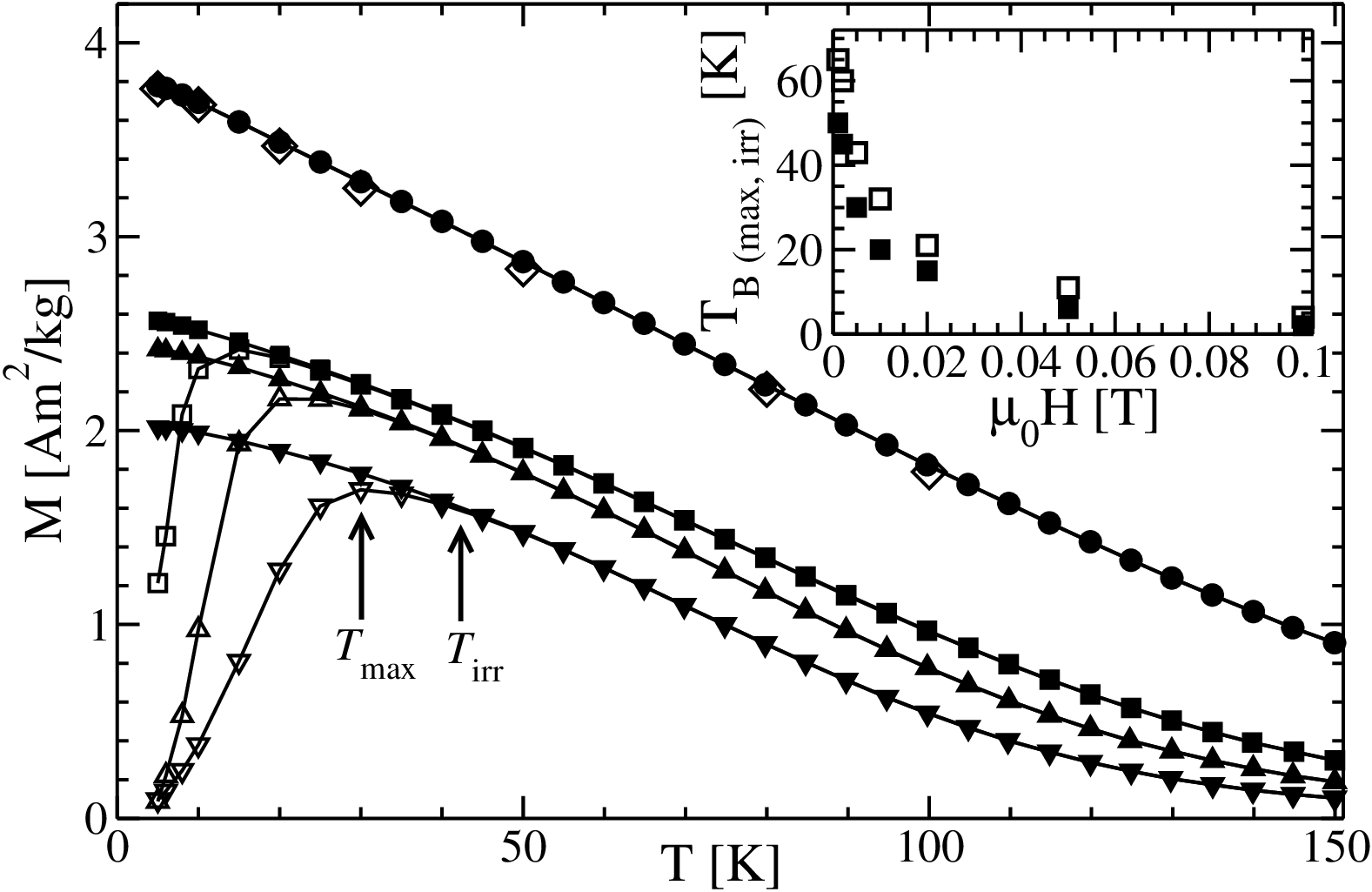}}
\caption{Zero-field-cooled (ZFC, hollow symbols) and field-cooled (FC, full symbols) $M(T)$ curves in different applied magnetic fields $\mu _{0}H $: 0.005T ($\fulltriangledown $), 0.01T ($\fulltriangle $), 0.02T ($\fullsquare $), 0.1T ($\fullcircle $) and from $M(H)$ curves ($\opendiamond $); lines are eye-guides. Inset: The dependence of blocking temperatures $T_{max} (\fullsquare )$ and $T_{irr} (\opensquare )$ on applied magnetic field $\mu _{0}H$. }
\label{zfcfc}
\end{figure}

The temperature dependence of magnetization $M(T)$ is measured in ZFC and FC modes for several temperatures and some of them are shown in \fref{zfcfc}. The well-pronounced ZFC-FC splitting points to the blocking of magnetic moment of the sample below $T_{B}$. It has been observed that $T_{B}$ lowers as the applied magnetic field $H$ increases. 

\begin{figure}[t]
\centerline{\includegraphics[width=9cm,clip]{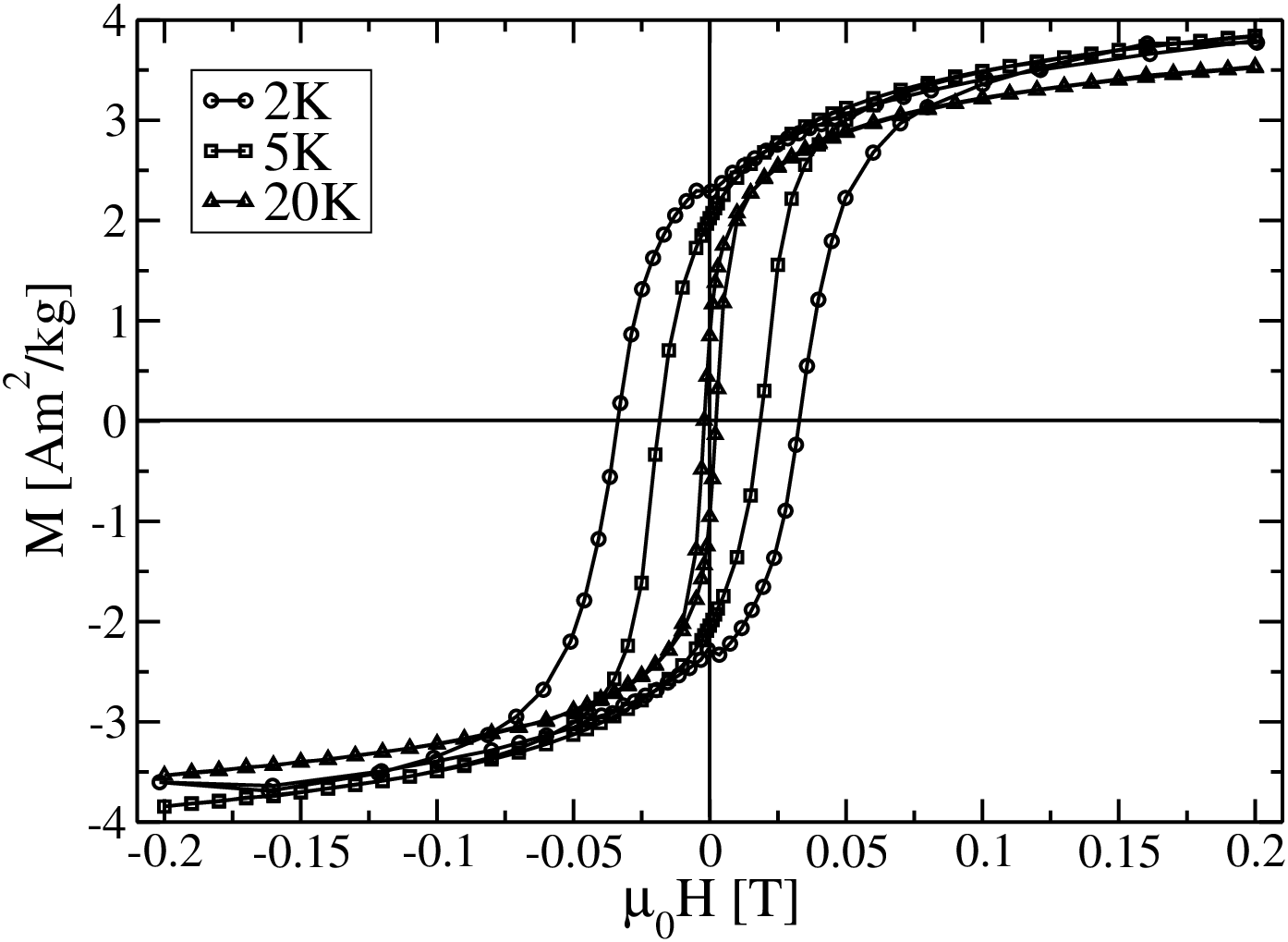}}
\caption{Magnetic hysteresis loops $M(H)$ at some temperatures with maximum applied field of 0.2T. Lines are eyeguides.}
\label{hyst}
\end{figure}

Some of measured hysteresis loops of the sample are presented in \fref{hyst}. They get narrower as the temperature increases, becoming almost reversible above approximately 30K. Obviously, the irreversibility is destroyed by thermal effects. For all measured temperatures the irreversibility appears below the field of 0.14T, so that the maximum field of 0.2T was high enough to study the hysteretic properties.

\begin{figure}[b]
\centerline{\includegraphics[width=9cm,clip]{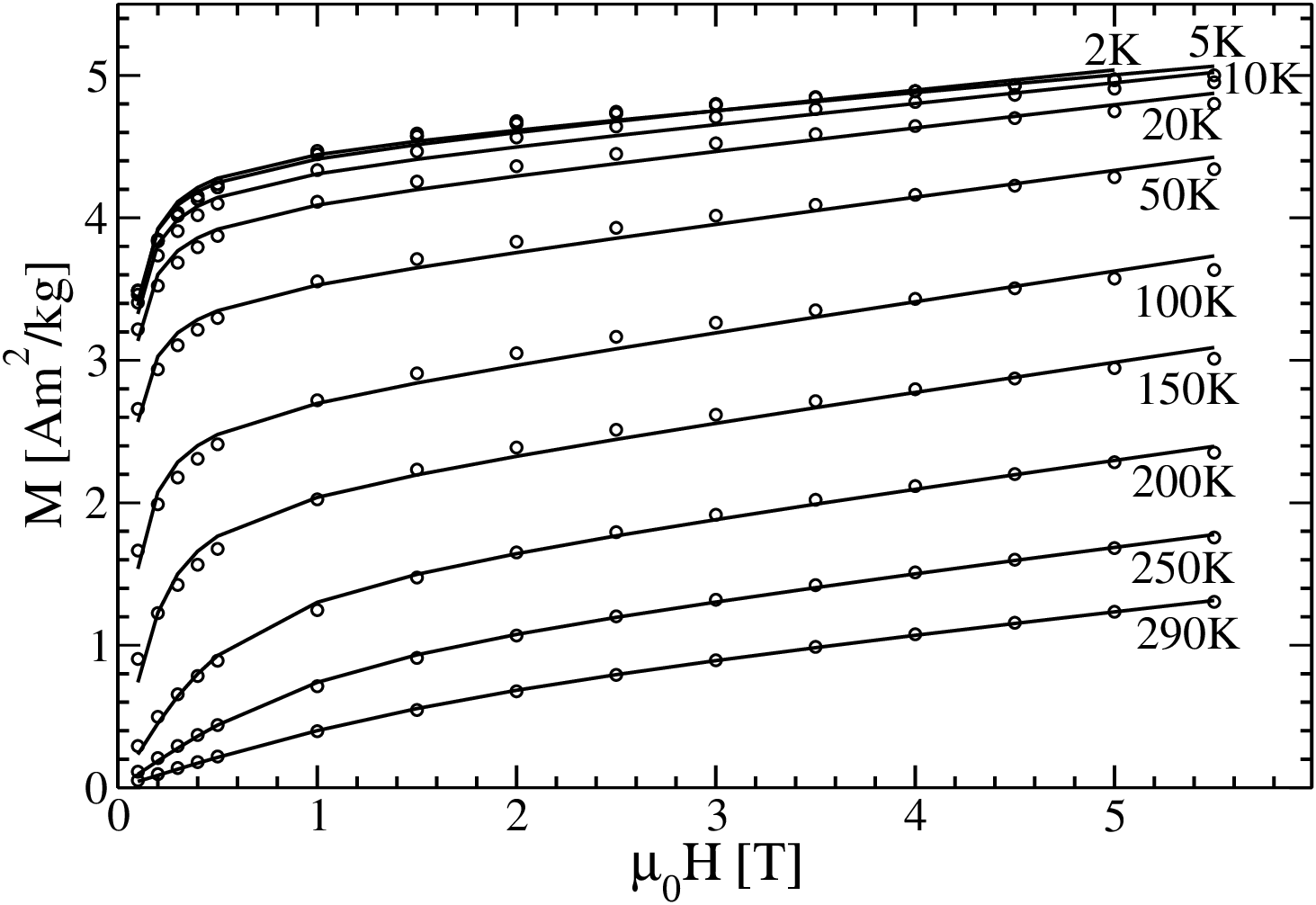}}
\caption{Dependence of magnetization $M$ of the sample on applied magnetic field $H$ for different temperatures. Lines are fitted functions (see equation \eref{langev} and \fref{langfit}).}
\label{m-h}
\end{figure}

Presented hysteresis loops are still far from saturation. $M(H)$ dependence is measured up to maximum possible field of 5.5T for different temperatures and plotted in \fref{m-h}. The saturation magnetization is unreachable still at 5.5T and the lack of $H/T$ scaling tells that the pure Curie-Brillouin-Langevin approach is not applicable. The $M(H)$ dependence is nonlinear even for small magnetic fields of 0.002T and up to room temperature. Documented paramagnetic susceptibility of hafnium is 0.42$\cdot 10^{-2}$J/T$^{2}$kg at room temperature with a weak temperature dependence (0.4$\cdot 10^{-2}$J/T$^{2}$kg at 77K and 0.46$\cdot 10^{-2}$J/T$^{2}$kg at 4.2K) \cite{hafnium}. Recalculated, it amounts only $\approx $0,5\% of mass magnetization in our sample at 100K and 5.5T, assuming it unjustifiably as an independent additive contribution.

\begin{figure}[t]
\centerline{\includegraphics[width=9cm,clip]{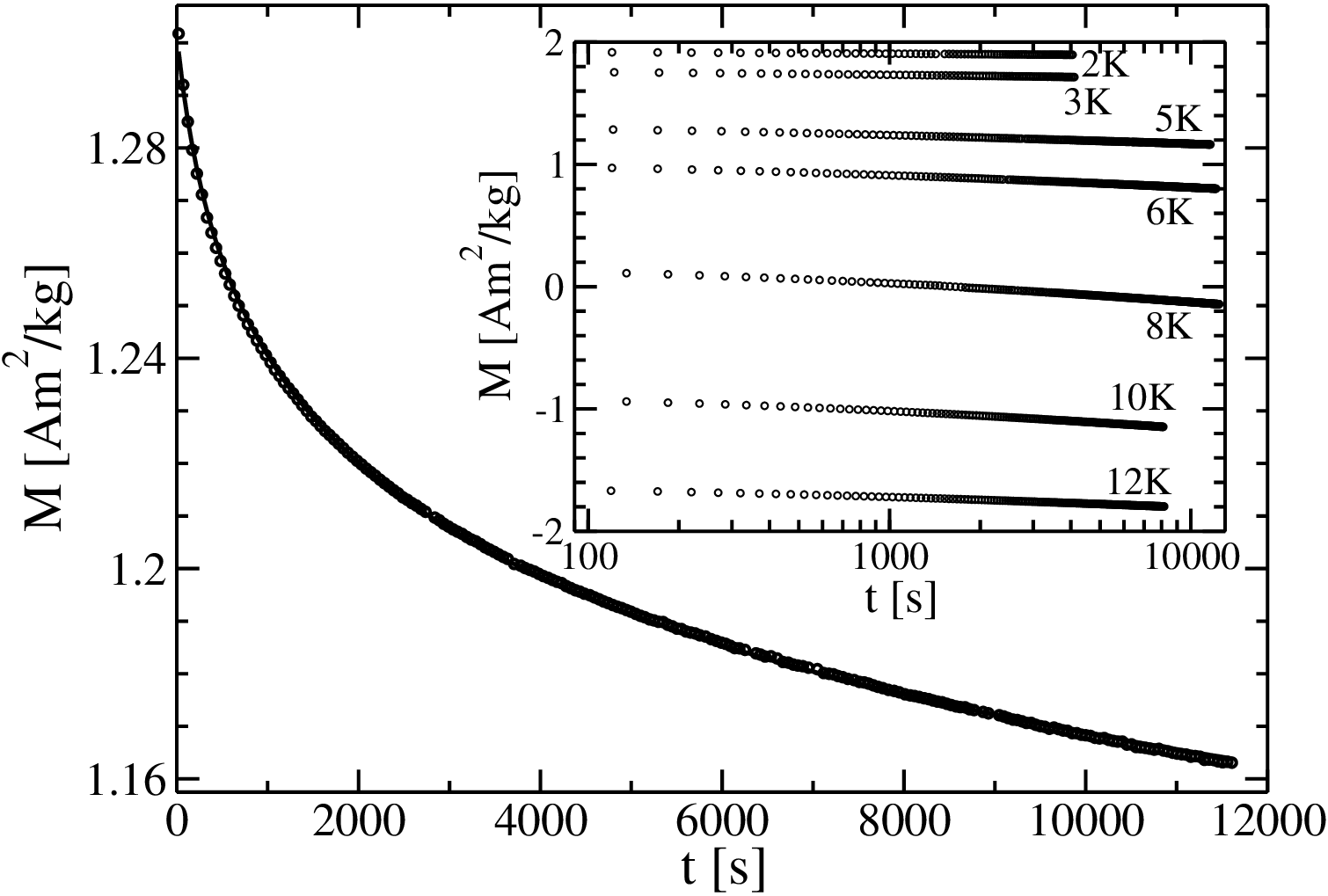}}
\caption{Relaxation of magnetization at 5K with logarithm fitting curve. Inset: Relaxation of magnetization at some temperatures on a logarithmic time-scale.}
\label{relax}
\end{figure}

The blocking of magnetic moments below $T_{B}$ results with the slow and measurable relaxation of magnetization presented in \fref{relax} for some of the measured temperatures. All of measured relaxation data for more than 40 different temperatures between 1.8K and 25K are logarithmic in time. 

AC susceptibility has been recognized as a powerfull tool to investigate the properties of the systems which show superparamagnetic and/or glassy behaviour. Especially, the measurements of the third harmonic ($\chi _{3}$) have been used to differentiate between these two systems which show very similar behaviour in the first harmonic ($\chi _{1}$) \cite{Bitoh1993}. It has been theoretically predicted that the spin-glass system should show a divergence of the third harmonic \cite{Fujiki1981} at the transition temperature $T_g$ which has been confirmed experimentally \cite{Bitoh1993,Bajpai2001}. On the other hand, the superparamagnetic systems show nondiverging peak in $\chi _{3}$ around the blocking temperature $T_B$ with a $T^{-3}$ dependence above $T_B$
\begin{figure}[b]
\centerline{\includegraphics[width=9cm,clip]{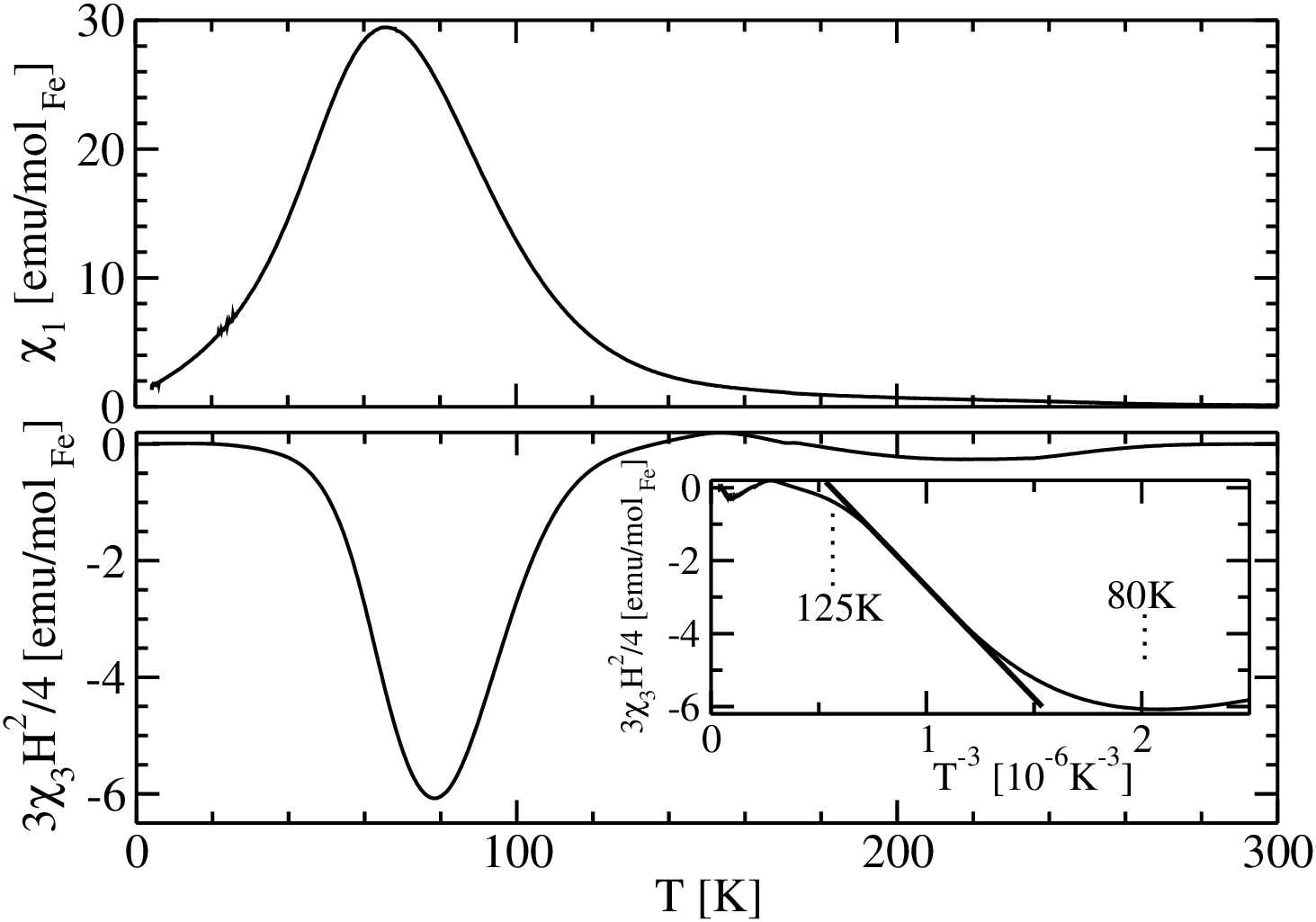}}
\caption{Temperature dependence of linear magnetic susceptibility $\chi _{1}$ (up) and nonlinear susceptibility $\chi _{3}$ transformed to $4/3\cdot \chi _{3}H^{2}$ (down) measured in 1mT and 990Hz. Inset shows the dependence of nonlinear susceptibility on $T^{-3}$.}
\label{acsusc}
\end{figure}
\cite{Bitoh1993,Bajpai2000}, in accordance with the Wohlfarth's blocking model \cite{Wohlfarth1979}. In \fref{acsusc} we present the results of the measurements of the first and the third harmonic of the ac susceptibility as a function of temperature. $\chi _{1}$ shows a round maximum at 66K and a monotonous decrease at higher temperatures. $\chi _{3}$ is zero below 20K, shows a minimum at 79K, a small broad feature centered around 220K and eventually goes to zero for $T>300$K.

Analysis of the temperature dependence of $\chi _{3}$ around the relatively broad minimum showed nondiverging behaviour, exluding the possibility that a spin-glass order is present in Hf$_{57}$Fe$_{43}$. In the inset we show $\chi _{3}$ {\it vs.} $T^{-3}$ dependence, predicted to be a straight line in the case of superparamagnets \cite{Wohlfarth1979}. The range in which a linear behaviour is observed is rather small ($\sim 10$ K). Recently, a similarly small interval has been observed in Li$_{0.5}$Ni$_{0.5}$O system \cite{Bajpai2000} and it has been attributed to a small difference between the blocking temperature $T_B$ and the intra-particle's spin-correlation temperature, which is much larger in conventional superparamagnetic systems. Also, it is possible that the deviation from linearity comes from the superposition of susceptibilities of superparamagnetic clusters which have different blocking temperatures.

Small broad feature in $\chi _{3}(T)$ graph is located around 220K. We observe that neither $\chi _{1}$ nor magnetization show visible deviations in that temperature range so we believe that it does not influence the main results presented in this paper. We will address this question in the future.

\section{Analysis and discussion}\label{discussion}

Performed magnetic characterisation alone gives many useful details about magnetic properties and processes in amorphous Hf$_{57}$Fe$_{43}$ magnetic alloy. In general, there is a lack of precise physical models of such kind of magnetic alloys and the descriptions are more phenomenological. Results presented in \sref{results} fit very well within the frame of thermal activation of blocked magnetic moments of superparamagnetic clusters, which is assisted by applied magnetic field. Now, different parameters derived from the presented raw data will be analysed within this frame in contexts of some existing models.

\subsection{Blocking of magnetization}

Taking for the origin of blocking the magnetic anisotropy barrier of height $U$, the relaxation time $\tau $ of the magnetic moment of the particle/cluster at temperature $T$ is determined by the activation law \cite{fluctdom,mqtmm}
\begin{equation}
\tau = \tau _{0} \cdot \exp (U/k_{B}T) 
\label{Arrh}
\end{equation}
where $\tau _{0}$ is of the order $10^{-9}-10^{-11}$s \cite{mqtmm}. At $T=T_{B}$ the relaxation time becomes equal to the time of measurement of one point $\tau =\tau _{exp}$, which is about 100s in our experiment. Using the blocking temperature for applied magnetic field 0.01T of $T_{max}=$20K and $T_{irr}=$30K and taking \eref{Arrh} it follows $U \approx 27 k_{B} T_{max} = 7.3 \cdot 10^{-21}$J and $U \approx 27 k_{B} T_{irr} = 1.1 \cdot 10^{-20}$J, corresponding to barrier heights responsible for blocking of moments below 20K and 30K, respectively, in magnetic field 0.01T. It is taken $\tau _{0}=10^{-10}$s \cite{mqtmm}, knowing that there are many difficulties \cite{tau0}, but fortunately the estimation of barrier height does not depend significantly on the chosen values for $\tau $ and $\tau _{0}$. 

In applied magnetic field $H$ the barrier height which prevents the escape of magnetic moment is reduced as $U = K V (1-\mu H/2KV)^{2}$, where $K$ is anisotropy energy density, $V$ is the volume of the cluster and $\mu $ is magnetic moment of the cluster. In our case, there is no reason for clusters to be of single size and have unique barrier heights, nor to be equally oriented. Therefore, $T_{B}(H)=(K V)/[ k_{B} \ln (\tau _{exp}/\tau _{0}) ] \cdot [ 1-(\mu H)/(2 K V)] ^{2}$ derived for ensemble of single sized equally oriented nanoparticles \cite{superparamagnetism} is not expected to fit the extracted $T_{max,irr}(H)$ dependence. Contrary, it is known to describe correctly the blocking temperature in single molecule magnet Mn$_{12}$-acetate \cite{mn12}, where all magnetic units are equal.

Also, the measured $T_{B}(H)$ can not be described by $H^{-2/3}$ nor any other similar power law dependence that is characteristic for some spin-glass and/or superparamagnetic systems \cite{TBHpower}. Pure exponentials are not suitable, too. However, measured FC curve with broad peak is not characteristic for spin-glass behaviour \cite{mydo}, but corresponds more likely to distributed blocked superparamagnetic clusters.

For measurement in 0.1T there is no ZFC-FC splitting down to the lowest measured temperature of 5K, that is consistent with hysteresis loops from \fref{hyst}. Also, it is shown in \fref{zfcfc} that the ZFC and FC values of magnetization for 0.1T are equal to the values taken from the hysteresis curves. This overlap says that in case where the applied field destroys the irreversibility, the history of magnetising process does not play a role.

\subsection{Magnetic hysteresis}

Question about the origin of the magnetic hystereticity in amorphous Hf$_{57}$Fe$_{43}$ is to be answered by looking at the slow relaxation of superparamagnetic units, indicated by the above presented results. The lowering of $T_{B}$ with the rising of magnetic field enlightens the origin of hysteretic irreversibility: the applied magnetic field changes the barriers established by anisotropy and helps the magnetic moments to overcome the barriers, so that the system should be on lower temperature in order that the higher number of clusters stays on the same side of barrier for a considerable amount of time. From the other side, increasing the temperature makes the loops get narrower (\fref{hyst}) indicating that the field needed to overcome the barrier is smaller because the moments already have higher thermal energy. Mechanism of magnetic hysteresis in case of heterogeneous alloys was analysed precisely by Stoner and Wohlfarth \cite{stonwohl}, but without any reference to temperature dependent dynamics. 

\begin{figure}[t]
\centerline{\includegraphics[width=9cm,clip]{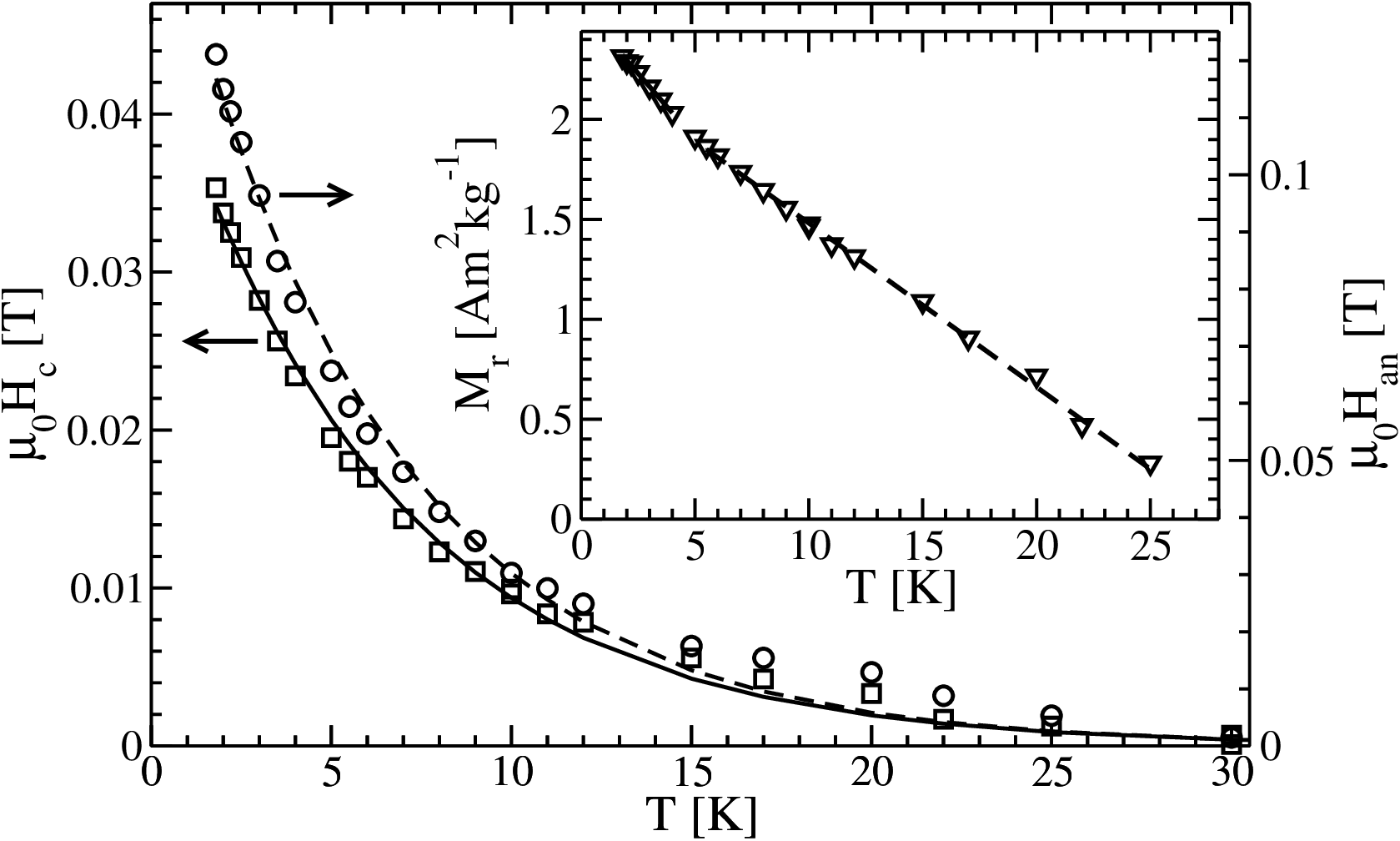}}
\caption{Temperature dependence of coercive field $H_{c}$ ($\opensquare $), anisotropy field $H_{an}$ ($\opencircle $) and remanent magnetization $M_{r}$ ($\opentriangledown $). Lines represent fitting curves.}
\label{hchanmr}
\end{figure}

The measured temperature dependence of coercive field $H_{c}$ is shown in \fref{hchanmr}. The data are fitted very well by exponential function $H_{c}(T)=H_{c0}\exp (-\alpha T)$. Best agreement is achieved for $\mu _{0}H_{c0}$ = 0.0455T and $\alpha $ = 0.158K$^{-1}$. 

When applied field $H$ is high enough to decrease the previously discussed energy barriers to $\approx $25$k_{B}T$, the reversal process can be thermally activated within the time of one measurement \cite{superparamagnetism}. The lack of consequently proposed dependence $ H_{c}=2 K V/\mu \cdot \{ 1 - [ (k_{B} T \ln (\tau_{exp}/\tau _{0}))/(K V) ]^{1/2} \} $ in describing the data shows again that the anisotropy barriers in our ensemble of magnetic clusters are not uniform.

The exponential dependence $H_{c}(T)$ describes very well the behaviour in other different systems. In a magnetic garnet film the coercivity was connected to the temperature dependence of anisotropy and simple exponential model was an ideal fit \cite{softcoerc}. The rare-earth-transition-metal random magnet (FeSm) exhibits also this kind of $H_{c}(T)$ dependence \cite{randmagrelax}. In such systems characterised by the strong ferromagnetic exchange and random magnetic anisotropy the atomic magnetic moments are correlated on a small scale, while on a large scale the magnetization rotates stochastically through the sample. There $H_{c}(T)$ is exponential for different exchange and anisotropy values \cite{randaniscoerc}. Exponential dependence was observed in FeZr amorphous alloy, Dy$_{60}$Fe$_{40}$, (Gd$_{1-x}$Tb$_{x}$)$_{2}$Cu, and also in many simulations and theoretical calculations \cite{hcexpmany}. 

In Hf$_{57}$Fe$_{43}$, the exponential dependence appears because of coercive field measures the difficulty of reversing the whole system of magnetic clusters which change their magnetic moments orientation over the anisotropy energy barriers by thermal relaxation. Our value of $\alpha $ is half of the value obtained in the random anisotropy model \cite{randaniscoerc} where the coercivity was investigated in the limited range of the anisotropy to exchange ratio and the decrease of $\alpha $ with decrease of exchange was observed. According to that study, our value of $\alpha $ points to the negligible exchange interaction between the units. The most complete simulation of coercivity in single-domain particle system \cite{coercivcomplete} includes the contributions from distributed blocked particles and superparamagnetic particles. Accordingly, our results point to the relatively broad distribution of cluster sizes. Additionally, with the power law, predicted for the disordered spins on nanoparticle surfaces \cite{surfacepower} and randomly oriented particles under the thermal influence \cite{hcpower}, it is not possible to fit the data. 

The temperature dependence of remanence in random anisotropy model was found to be exponential, too \cite{remanexp}. In our measurements $M_{r}(T)$ is linear in two regions (shown in \fref{hchanmr}). At lower temperatures (below 5K) the change of remanence with temperature is --0.132Am$^{2}$/kgK, at higher temperatures (5-25K) it is --0.082Am$^{2}$/kgK, and above 25K the remanence fluctuates at small values. Here, slower decrease happens for longer duration of hysteresis loop measurement. This qualitative correlation is argued by the design of experiment where the measurement was performed faster below  the 4.2K (liquid helium). Generally, the memory related to remanence lasts longer with lower temperature. It is reasonable that the system memorises the state more intensive when the change of the applied field is faster, because the system has no time to come closer to the new equilibrium. Altogether, the slower change of remanence is observed when the change of field is slower, due to the lower ability of memory alone. This explanation fits very well within the frame of dynamical hysteresis caused by thermal activation over the anisotropy barrier. Linear dependence of remanent magnetization on temperature was found in simulation of magnetic processes in amorphous alloy with nanometre sized magnetic clusters with random distribution of orientation \cite{remanlin}. This is applicable also for random distribution over sizes, including our system.

Anisotropy field $H_{an}$ depends on temperature as shown in \fref{hchanmr}. $H_{an}$ is obtained from hysteresis loops as the field above which the loops become reversible. This is reasonable in systems where the anisotropy axes are distributed in all directions, which is expected for our amorphous material. The clusters having perpendicularly oriented anisotropy axes with respect to magnetic field contribute weakly to the hysteretic irreversibility, but the clusters with axis oriented in the direction of applied magnetic field define the highest field at which irreversibility exists. Again, the exponential function $H_{an}(T)=H_{an0}\exp (-\beta T)$ (like in \cite{softcoerc}) was applied giving the parameters $\mu _{0}H_{an0}$ = 0.157T and $\beta $ = 0.165K$^{-1}$. The slight departure of the anisotropy field data from the exponential curve at temperatures above 15K, as well as for coercive field data, may be caused by a small amount of large clusters or by a very weak interaction between clusters which contribute with narrow hysteresis loops.

The measured ratio $H_{c}/H_{an}\approx 0.3$ in the temperature interval 2-20K (it decreases slightly above 20K) is in agreement with a value obtained for a system of randomly oriented single domain particles with a relatively broad distribution over sizes \cite{coercsupblock}.

Lowering of $H_{c}$, $M_{r}$ and $H_{an}$ with increasing the temperature appears because the system has more and more energy to turn the moments of the clusters over the barriers, so that the smaller magnetic field is needed to reverse the direction of magnetization, the system is less able to keep the moments blocked and the lower field is needed to make the system reversible.

\subsection{Relaxation of magnetization}

\begin{figure}[t]
\centerline{\includegraphics[width=9cm,clip]{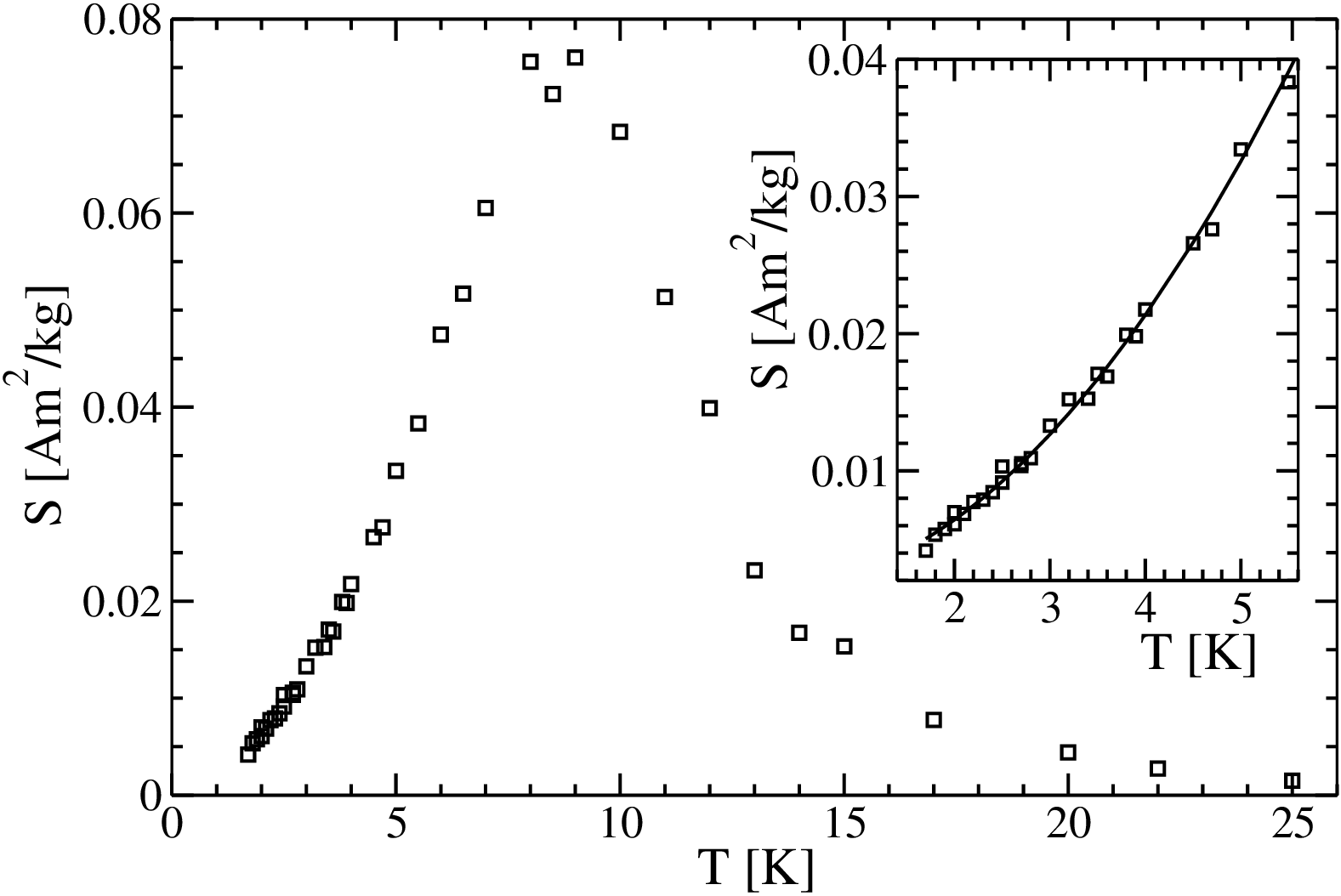}}
\caption{Temperature dependence of magnetic viscosity $S$ of the sample. Inset: low temperature region with regression curve.}
\label{visco}
\end{figure}

After the change of direction of applied magnetic field the system goes very quickly to the initial value of magnetization. This part of the fast magnetic relaxation is not measurable using our experimental device which has time resolution of $\sim $1min. From initial value which depends on temperature the sample is relaxing slowly toward the equilibrium determined by magnetic field and temperature, which is practically close to the ZFC value of magnetization. When plotted with logarithmic time-scale, all the relaxation curves appear linear and the measured relaxation data were fitted very precisely by
\begin{equation}
M(t)=M_{0} - S\cdot \ln (t-t_{0}).
\label{mtfit}
\end{equation} 
$S$ is called magnetic viscosity or logarithmic relaxation rate, $M_{0}$ is initial magnetization and $t_{0}$ has the meaning of time from which the system started to relax slowly. The fitting results for $S$ at different temperatures are shown in \fref{visco}. On its lower temperature part $S(T)$ reflects the increase of the relaxation rate with temperature. Contrary, the decrease of $S$ at $T>9$K in accordance with the magnetic moments blocking hypothesis does not mean that the relaxation becomes slower, but just that a small part of moments remained to relax after the main part of the sample has been relaxed prior to taking any measurement at this temperature. Parameter $-t_{0}$ is mainly between 100s and 200s for all temperatures, which is somewhere during the superconducting coil recharging time and it is consistent with the assumed fast arrival to the initial magnetization before the measurement.

Some microscopic results can be obtained when interpreting that data in the model of activation over barriers. The logarithmic relaxation of magnetization in the ensemble of superparamagnetic units was derived using the notion of critical barrier height \cite{SkTCkT}. Ensemble of magnetic entities consists of characteristic magnetic moment units with corresponding energy barriers. From already discussed activation law \eref{Arrh} it follows that in the distribution of energy barriers there is a critical one, above which the moment of the cluster is stable on the time scale of the relaxation measurements. The clusters with lower barriers relaxed to their equilibrium determined by the applied magnetic field $H<0$ prior to taking the measurement. The clusters with higher barriers remain blocked in the original state determined by cooling field $H>0$. As the time passes, some additional moments jump over the barrier and the magnetization progresses toward the new equilibrium determined by applied field. Because of the impossibility to cover the whole range of exponentially distributed relaxation times with the measurement at one temperature within the finite time, it is necessary to look just the limited time interval, which is very narrow comparing to the time scale covered by relaxations over all of available barrier heights. Changing the temperature at which the relaxation is measured, we cover different barrier heights probed during experimental time-window. Similar approach was applied in \cite{barrierTlnt}.

We are aware that the cutting of overparametrisation inside the logarithmic function makes problem with units, but it is useful if carefully interpreted. There is also approach with other functional dependence on time \cite{bessel}, which shows that because of slow relaxation distributed over many time decades it is possible to use small time-window of many functions. Despite the objections against the use of logarithmic fit \cite{bessel}, it turned out to be successfull in other cases \cite{feptdistrib,rmaqtm}, so we used it.

Likewise, starting from the exponential relaxation of magnetic nanoparticles in ensemble with distributed size slightly different form of magnetic viscosity is presented in the textbook \cite{mqtmm}. For low temperature magnetic viscosity it was found $S\propto T^{2}$. The addition of constant term in $S = a T^{2} + b$ was necessary in order to fit our low temperature data in \fref{visco} for $T\le 5.5$K resulting with $a$=0.00124a.u. and $b$=0.00145a.u. Finite $b$ would eventually point to quantum tunnelling of magnetization possibility, but for this claim lower temeratures should be investigated. 
In Cu$_{x}$Fe$_{3-x}$O$_{4}$ nanoparticles \cite{cufeo} and many other, $S(T)$ was found linear and tunnelling would manifest as a plateau in $S(T)$ \cite{tunelnano}.

To extract the barrier heights from \fref{visco} a link between temperature and barrier height is needed. Using for $\tau $ in \eref{Arrh} the characteristic duration of the relaxation experiment $\tau \sim 1000$s it follows $U\approx 30k_{B}T$. Maximum of $S$ is achieved at $T=8.5$K, corresponding to $U=3.5 \cdot 10^{-21}$J. This is barrier height for the clusters with highest magnetization contribution in whole system. Above this value $S$ decreases abruptly up to $U=7 \cdot 10^{-21}$J showing that the amount of magnetic clusters with higher barriers decreases abruptly, too. Eventually, there is a small amount of clusters with higher barriers, up to $U= 10^{-20}$J, and still higher barriers were not probed by our experiment. The lowest measured temperature in $S(T)$ corresponds to $U=7.4 \cdot 10^{-22}$J. Lower barriers were not probed.

Blocking temperature $T_{max}$ from \ref{zfcfc} corresponds to the highest significant contribution in the $S(T)$ dependence from \fref{visco} making it as the temperature at which almost all of the clusters have already relaxed to the equilibrium. On the other hand $T_{irr}$ corresponds to the value on this plot to which further small decrease in $S(T)$ is observed. This slight decrease amounts to small number of bigger moments which are stable much longer than the time-scale of experiment. Such a shape of $S(T)$ is the reason for $T_{max} < T_{irr}$. Also, shape of $S(T)$ is the occasion to exclude the spin-glass possibility as a cooperative random freezing. The ZFC curve starts to decrease broadly above $T_{max}$ because of fast fluctuations of the magnetization of smallest clusters, and at $T_{irr}$ when really all clusters relaxed it meets the FC curve (\fref{zfcfc}). So, the consistency between the magnetic viscosity and blocking temperatures is shown.

\subsection{Magnetic clusters sizes}

Magnetic cluster sizes can be obtained from barrier heights data if magnetic anisotropy density $K$ is known. $K$ is approximately calculated from hysteresis loops using 
\begin{equation}
K=\mu _{0}H_{an}M_{s}/2
\end{equation}
according to the well-known magnetic anisotropy models \cite{mqtmm,stonwohl,rmaqtm}. The dependence of anisotropy field $H_{an}$ on temperature is shown in \fref{hchanmr}. The extrapolation to zero temperature is taken from the fit in order to exclude the thermal effects on anisotropy: $\mu _{0}H_{an0}=$0.157T. 
The saturation magnetization $M_{s}$ is extracted from the high field magnetization variation. The data above 2T of $M(H)$ curve for the lowest measured temperature of 2K is fitted with the usual expression 
\begin{equation}
M=M_{s}\cdot \left( 1 - \case{4 K^{2}}{15 M_{s}^{2}H^{2}} \right) + \chi _{m} H  
\end{equation}
based on model from \cite{msat}. The paramagnetic susceptibility $\chi _{m}$ of matrix becomes somewhat bigger than $\chi _{Hf}$ when appropriately scaled taking the mass contribution $w$(Hf)=0.819. From obtained $M_{s}=4.6$Am$^{2}$/kg and the mass density of Hf$_{57}$Fe$_{43}$ $\rho \approx 1.2\cdot 10^{4}$kg/m$^{3}$ \cite{density}, it follows the volume saturation magnetization $M_{s}^{V}=5.6\cdot 10^{4}$A/m. This should be divided by iron mass contribution $w(Fe)=0.191$ because the paramagnetism of hafnium is much weaker than the contribution of iron. It follows $K=2.1\cdot 10^{4}$J/m$^{3}$. The same fitting gives also directly parameter $K=3.0\cdot 10^{4}$J/m$^{3}$. This value is used because nearly the same is obtained by the magnetization-area method \cite{msat} applied to our data. The difference between two results is attributed to the influence of third free fitting parameter $\chi _{m}$. The biggest countable volume of the magnetic clusters are now $V=U/K=230$nm$^{3}$, that would correspond to the sphere of a diameter of 7.4nm (or a cube with 5.9nm side). For the clusters with highest contribution to magnetization (maximum at the plot in \fref{visco}) the volume is $V=U/K=110 $nm$^{3}$, that would correspond to the diameter of a sphere of 6.1nm (or a cube of 4.9nm side). For the cluster with lowest measured barrier the volume is $V=U/K=25$nm$^{3}$, that would correspond to the diameter of a sphere of 3.6nm (or a cube with 2.9nm side). We can say nothing about smaller clusters concerning our measurements. The lower temperatures should be probed for them, or the measuring technique of higher frequencies used. Nevertheless, their contribution to the magnetization is very small. There is an estimated number of iron atoms based on the volume of the clusters and mass density in magnetic clusters which is mainly from about 500 to 5000, with a small amount of bigger clusters of up to $\sim 10000$ iron atoms. 

$M(H)$ curve measured at 100K can be described very well with rectangular distribution over volumes of clusters. The magnetization of each cluster is given by Langevin function $L(\mu H/k_{B}T)$, where $L(x)=1/\tanh (x) - 1/x$. Flat distribution over magnetic moments is used and the matrix paramagnetic term $\chi _{m}\cdot H$ with known $\chi _{m}$ from previous measurement is added taking care about the mass contributions of iron and hafnium. It is not very suitable to fit final function to the small number of data points because of many fitting parameters, but also because of physical reason: the shape of $M(H)$ does not depend very much on the distribution function, but more on the mean value of magnetic moment, as is well known \cite{coercivcomplete}. Instead, bare plotting shows good agreement between the curve and the measured points using $\mu _{min}=800\mu _{B}$ and $\mu _{max}=5000\mu _{B}$, that is consistent with number of iron atoms obtained from relaxation analysis. Significantly different values of $\mu _{min}$ and $\mu _{max}$ give the significant deviation between the calculated and measured $M(H)$ curves. This points to the consistency of the established picture of Hf$_{57}$Fe$_{43}$ system. 

\begin{figure}[t]
\centerline{\includegraphics[width=8cm,clip]{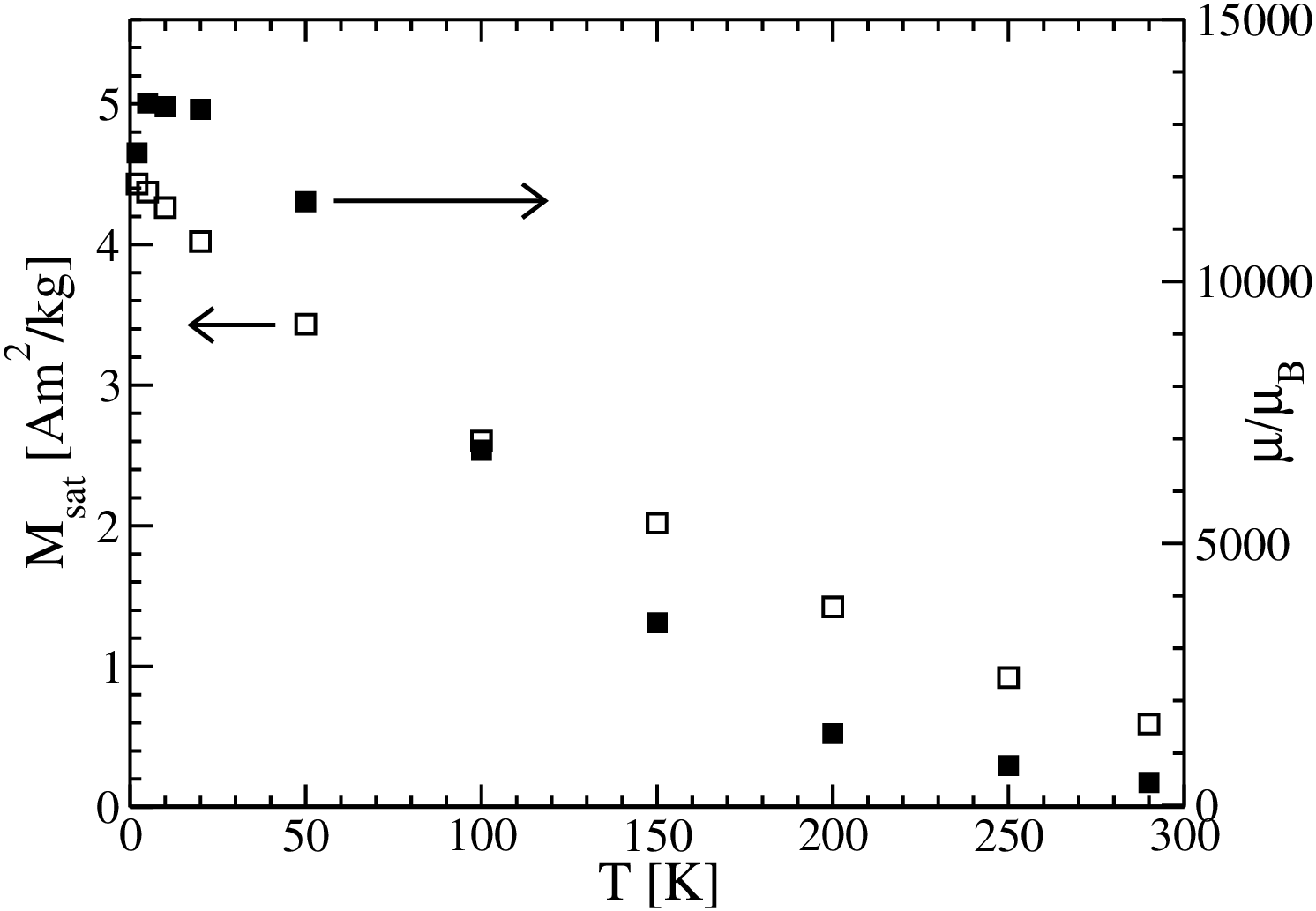}}
\caption{The temperature dependences of effective magnetic moment of the magnetic clusters $\mu $ ($\fullsquare $) and saturation of magnetization of clusters $M_{sat}$ ($\opensquare $) derived from $M(H)$ curves.}
\label{langfit}
\end{figure}

The simplification is performed using a superposition of the superparamagnetism of single sized clusters and paramagnetism of nonmagnetic matrix to describe $M(H)$ curves from \fref{m-h}. The fitting of
\begin{equation}
M(H)=M_{sat}\cdot L(\case{\mu H}{k_{B}T}) + \chi _{m}\cdot H
\label{langev}
\end{equation}
to the measured $M(H)$ gives the temperature dependence of effective magnetic moment $\mu $ and saturation magnetization of clusters $M_{sat}$ shown in \fref{langfit}. Third parameter, the matrix paramagnetic susceptibility $\chi _{m}$, stays between 0.12-0.20J/T$^{2}$kg. It is much bigger than published data for pure hafnium \cite{crangle,hafnium} that is in accordance with fast increase of paramagnetic susceptibility with iron concentration in Hf-Fe \cite{HfFeRistic} and Zr-Fe \cite{FeZr} systems. Our value of saturation magnetization agrees well with the values for Hf-Fe system studied in \cite{hffemicto} for higher iron concentrations. The deviation of fitted curves from measured $M(H)$ in \fref{m-h} becomes more expressed as temperature decreases. Obviously, the true characteristic magnetic moments below 100K are smaller than shown in \fref{langfit}, and they would correspond more nearly to values obtained from relaxation analysis. Nevertheless, the parameters shown in \fref{langfit} are indicative for the behaviour of material. The decrease of magnetization saturation with temperature can be understood as destroying of magnetic ordering of clusters, that is in agreement with the decrease of magnetic moment.

\subsection{Additional remarks}

As shown, the presented data are explained very well with the notion of magnetic clusters. This is supported by investigation of other concentrations of iron in Hf$_{100-x}$Fe$_{x}$ amorphous system, where the long range magnetic order for higher concentration of iron ($x\ge 50$) \cite{HfFe} and no magnetic ordering for lower iron concentration ($x \le 40$) is observed \cite{HfFeRistic}. The iron concentration in Hf$_{57}$Fe$_{43}$ is well above the percolation threshold, but it seems that magnetic clustering is preferred over the long range ordering. Reasons for this should be explored by other experimental techniques. Maybe a good reference would be one indirect evidence of non-homogeneous atomic co-ordination in Hf$_{57}$Fe$_{43}$ metallic glass \cite{glasses}.

At the end, few remarks can throw more light on the presented system and give impulse for further investigation.

The temperature independence region in $S(T)$ was not observed down to the lowest measured temperatures (1.8K). The extrapolation of $S(T)$ to zero temperature gives the non-zero relaxation rate. It is still the question if such signs of quantum tunnelling of magnetization can be observed in this kind of material at lower temperatures, as it was seen in some random anisotropy magnets \cite{rmaqtm,tunnelalloy} and ensemble of magnetic nanoparticles \cite{particletunnel}. Theoretical predictions say that magnetization tunnelling could be observable below $\approx $0.1K \cite{particletunnel} in our case.

Additional measurement of $M(H)$ was performed so that every point is taken after the zero-field cooling. The perfect overlap of this $M(H)$ curve with "single-shot" $M(H)$ measurement shows that every point was measured in equilibrium in both cases, excluding thus the spin-glass freezing. Also, the memory effects are investigated in another way \cite{memorychaps}. A full hysteresis loop is cycled, then the field is reduced sweeping from the maximum positive field to negative field somewhere around the half of $H_{an}$, and after that the field is swept from this negative value toward the maximum positive field. The kink at the same positive field around the half of $H_{an}$ was not observed. This demonstrates that in Hf$_{57}$Fe$_{43}$ there is no memory effect connected with macroscopic number of the frustrated symmetric clusters in the spin-glass frame \cite{memorychaps}.

Hysteresis curves are not shifted after field cooling, excluding further the memory effect. Also, the exchange field between non-magnetic matrix and magnetic clusters is negligible. Finite exchange should induce some ordering of the cluster's surface layer which should couple with cluster's core magnetization, shifting thus the hysteresis loops if measured below the freezing temperature \cite{exchbias}. This is known as the exchange bias. Furthermore, the lack of the shift says that there is no disorder of the cluster's surface layer which was observed in some magnetic nanoparticle systems \cite{exchbiasexp} and that there is no antiferromagnetic shell around the core of the cluster. It is possible that the magnetic clusters are below the critical dimension for the onset of the exchange bias \cite{exchcrit}, or that freezing appears below 2K.

When considering the relaxation properties, one should have on mind that it is measured just for one applied magnetic field (0.01T). From the analysis of blocking temperature it follows that in zero field the characteristic barriers will be twice greater than the barriers when the field is 0.01T. The influence of magnetic field on the barrier heights will be studied in the future. Furthermore, presented measurements say that the high temperature investigation (from 30K to room temperature or above) would be another interesting topic.

This investigation gives also some hints for a better understanding of magnetism in similar but more investigated Zr-Fe system. It has been thought that below the critical concentration of iron the spin-glass behaviour appears \cite{FeZr}. Our work shows that the question about the possible blocking of superparamagnetic clusters under some conditions should be raised, too.

\section{Conclusion}\label{conclusion}

All magnetic measurements performed on binary magnetic/nonmagnetic amorphous alloy Hf$_{57}$Fe$_{43}$ point to its superparamagnetic behaviour and magnetic moment blocking of the clusters. Superparamagnetism is argued with concentration of iron which is a little below the critical threshold for the long range magnetic ordering, so that the finite magnetic clusters separated by nonmagnetic regions are expected. 

ZFC and FC curves, temperature dependent hysteresis loops, slow relaxation of magnetic moment and temperature dependence of magnetic viscosity all show that the magnetic clusters change the direction of their magnetic moment over the magnetic anisotropy barrier by thermal activation. The phenomenological description of the magnetic relaxation using the mentioned quantities and concepts provides a useful link between the experiment and microscopic model. Definitely, the nanometer sized magnetic clusters in Hf$_{57}$Fe$_{43}$ alloy are responsible for the observed magnetic processes. Their characteristic volume is estimated roughly to be 25-230nm$^{3}$ using just the magnetic measurements. 

The interaction between clusters and nonmagnetic matrix was not observed and the clusters show no memory effects connected to interaction with nonmagnetic environment or mutual interaction. Further investigation using magnetometric methods should be performed to describe more precisely the microscopic structure and properties of the magnetic clusters. 

Studied system is an excellent potential candidate for magnetization tunnelling investigation because it is free of surface disorder, exchange bias and other distorting phenomena, at least in the investigated temperature interval. This fundamental quantum process concerns the low-temperature behaviour and the measurements should be done at much lower temperatures than ours. On the other side, the high temperature behaviour is interesting for the investigation of development of magnetization disorder, which is obviously important for room-temperature properties of this class of materials.

\ack
We are grateful to Michael Reissner from Technical University in Vienna and Davor \v{C}apeta from University of Zagreb for very valuable discussions. Special thanks is devoted to Klara Bili\'{c} Me\v{s}tri\'{c} for reading the manuscript. This work was supported by the Croatian Ministry of Science, Education and Sports.

\end{document}